# Science Fiction Media Representations of Exoplanets: Portrayals of Changing Astronomical Discoveries

Authors: Emma Johanna Puranen, Emily Finer, Christiane Helling and V. Anne Smith

ABSTRACT: *Interest in science fiction's (SF's) potential science communication use is hindered by concerns about SF misrepresenting science. This study addresses these concerns by asking how SF media reflects scientific findings in exoplanet science. A database of SF exoplanets was analysed using a Bayesian network to find interconnected interactions between planetary characterisation features and literary data. Results reveal SF exoplanets designed after the discovery of real exoplanets are less Earth-like, providing statistical evidence that SF incorporates rapidly-evolving science. Understanding SF's portrayal of science is crucial for its potential use in science communication.*

**Context**

*Introduction*

Science and science fiction are not separate human endeavours. Scientists have been shown to demonstrate a strong connection to science fiction (SF) as fans and creators [Wright & Oman-Regan, 2018; Stepney, 2015]. A survey of 239 UK astronomers attending the 2022 UK National Astronomy Meeting provided statistical evidence that the majority were interested in SF and stated that SF influenced their career decisions [Stanway, 2022]. There is growing interest in analysing SF's utility for science communication [Joubert et al, 2019; Irani & Weitkamp, 2023] and the role SF plays in imagining futures for scientists and the public alike [Reinsborough, 2017]. Several studies analyse how SF affects its readership's understanding of science: Eichmeier et al investigated audience understanding of the human genome [2023], Lowe et al [2006] similarly investigated audience understanding of environmental science [2006], and Orthia interviewed *Doctor Who* viewers about how the show impacts their ideas about science [2019]. There are examples of scientists using SF for science communication: the European Astrobiology Institute recommends its science fiction anthology *Life Beyond Us* [Nováková, 2023] for educational use, several university courses use SF to teach science [Luokkala, 2014; Saunders et al, 2004], and scientists have presented their research at SF conventions in 'science programming' tracks [Childers et al, 2023]. However, SF has yet to be widely utilized as a science communication tool by full-time science communicators. Although interest in its potential science communication use has been established, and there is evidence that narratives are useful teaching tools for non-expert audiences, there are concerns about the inherent persuasiveness of narratives [Dahlstrom, 2014], and, significantly, concerns that SF misrepresents science [Childers et al, 2023; Lowe, 2006]. Given the interest and potential concerns about the accuracy of SF content, an important step in current scholarship is to examine not just audience reactions to SF, but the representations of science in the content of SF itself.

SF involving exoplanets—planets outside our solar system—presents us with an appropriate corpus for analysing the presentation of scientific discoveries in SF. This is because exoplanets are a very common setting in SF stories, but real exoplanets were only confirmed to exist by astronomers with the 1995 discovery of the first exoplanet orbiting a sun-like star, which completely changed scientific understanding of exoplanets and solar system formation [Mayor & Queloz, 1995]. While data about real exoplanets have only been available for the past three decades, imagined exoplanets have featured in SF for centuries. The example of exoplanets offers the opportunity to investigate changing representations of science in SF, which aids in determining its use for science communication.

This study proposes a new methodology for analysing the portrayal of science in SF. We take inspiration from the handful of literary scholars who have argued for the scaling up of literary studies to analyse trends and genres rather than individual works and authors. We employ a Bayesian network as a statistical methodology that allows us to map interconnected variables of fictional exoplanet characteristics in our database of SF exoplanets to identify larger contexts and indirect connections. We relate the conclusions we draw from this study of the science in SF to the potential use of SF in science communication.

*Science fiction*

Science fiction is defined broadly as stories featuring "cognition" and "estrangement", that is, changes from reality presented in an empirical, rational manner within the fictional universe of the story [Suvin, 1979]. SF authors emphasize the genre's strong relationship with science in their own definitions, with Asimov defining SF as "literature which deals with the reaction of human beings to changes in science and technology" [1975]. SF is a significant transnational cultural phenomenon. SF films consistently take 10-20% of the box office market share [Nash Information Services, LLC, n.d.]. SF books like Cixin Liu's *The Three-Body Problem* and Andy Weir's *The Martian*, and video game franchises like *Mass Effect* sell millions of copies [McGregor, 2020; Alter, 2017; Radić, 2021]. SF is not merely a literary genre—SF fans read, watch, listen to, and play SF stories, and this is reflected in the source works of the fictional exoplanets in this study.

Science fiction and science are widely accepted to be interdependent phenomena, but it is vital to characterise the relationship more attentively. Ursula Le Guin [1969] famously said: "science fiction is not predictive, it is descriptive" [p. xiv], that is, it reflects what it sees in the contemporary world in which it is written, including the science of the day. However, recent evidence argues for a less passive role for SF in society and science. There is evidence that SF attracts readers to choose scientific careers [Wright and Oman-Reagan, 2018; Stanway, 2022]. Estébanez Camarena argues that "Science fiction inspires scientists to realize the visions that it expresses", citing examples from SF that became science fact, such as Jules Verne's rocket in his 1865 novel *From the Earth to the Moon* and the space station in the 1968 film *2001: A Space Odyssey* [2020, p. 130].

Scientists sometimes instrumentalise SF to communicate science. SF can offer a common knowledge base: we, the authors, frequently refer to a fictional planet (Tatooine) in our public engagement work to help audiences understand the term exoplanet. Popular science books like Joshua Winn's *The Little Book of Exoplanets* and Jaime Green's *The Possibility of Life* extensively cite fictional exoplanets as examples. University courses have been developed using SF to teach scientific concepts to audiences largely composed of humanities students, building on people's passion for the genre to instil an interest in learning science [Luokkala, 2014; Saunders et al, 2004]. SF conventions frequently feature science programming tracks where scientists present their research to the public and relate it to SF [Childers et al, 2023]. Hypotheses about SF's role in the public's exposure to scientific concepts are beginning to be supported by evidence-based research, as in Eichmeier [2023] which demonstrates that the consumption of SF influences audience perceptions of human genome editing. SF's potential beyond the communication of information, as a source of communication methods such as storytelling, or scenario-building to explore the societal consequences of scientific advances, are also increasingly noticed [Joubert et al, 2019; Irani & Weitkamp, 2023]. Scholars are considering what role SF might play in science communication, beyond these current examples of use by scientists [Dahlstrom, 2014; Reinsborough, 2017].

If an active role is to be assigned to SF as a driver and communicator of science, we must consider how best to measure the influence of science on SF and its presentation therein. There are some strategies available. We could focus exclusively on SF by scientists, for example Carl Sagan, Joan Slonczewski, or Isaac Asimov. Another method is to collect acknowledgements by authors in works of SF: where present, these may include lists of published sources, names of scientists and science advisers. A recently-published SF collection suggests such a new and more egalitarian model of co-authorship by pairs of scientists and creative writers who reflect on their joint process in the volume [Puranen, 2021]. Some secondary sources do exist, eg. Lawrence M. Krauss's *The Physics of Star Trek,* but these typically focus on a limited canon of popular works. However, in addition to being essentially subjective, none of this data can be collected on a large scale or assessed over time to determine if science fiction changes as scientific knowledge expands—a question to be answered if SF is to be used for science communication.

> *Research Question 1 (RQ1): How can **we study whether** science fiction reflects scientific discoveries?*

*Exoplanets*

To investigate if SF reflects scientific discovery, we need a scientific field that is well-represented in SF and has recently seen huge growth of new knowledge. Long-theorised but relatively recently observationally confirmed, exoplanet science fulfils these requirements. In SF, exoplanets have been frequent settings for stories since the early 20th century, and sporadically even before.

In science, since the 1995 discovery of 51 Pegasi b, a planet orbiting a sunlike star [Mayor & Queloz, 1995], over 5000 exoplanets have been discovered [NASA Exoplanet Science Institute, 2023]. The long time gap between theory and observation was due to the difficulty of exoplanet detection. Exoplanets are dim and challenging to observe; they are usually detected indirectly through observations of their host stars. Two common methods are *radial velocity*, a spectroscopic method which looks for Doppler shifts in a star's spectrum that indicates the star is affected by an exoplanet's gravity, and the *transit method*, which looks for a decrease in the stellar flux we receive from a star as its exoplanet passes between it and the Earth [Fischer et al, 2014]. Depending on the detection method, astronomers can determine mass, radius, and length of year of these distant worlds [NASA Exoplanet Science Institute, 2023].

In recent decades, the discovery of exoplanets has completely changed models of solar system formation and planet categories. For example, it had been thought that gas giants did not orbit close to their stars, but the first discovery of an exoplanet around a sunlike star showed that such "hot Jupiters" exist [Mayor & Queloz, 1995]. It was not until 2009 that the French CoRoT (Convection, Rotation et Transits planétaires) mission discovered the first rocky exoplanet [Léger et al, 2009]. Exoplanet science has gone from a niche field mostly focused on achieving detection to a major field within astronomy. It was chosen as a priority in the 2020 NASA decadal survey, a paper outlining upcoming priorities of the American astronomical community [National Academies of Sciences, Engineering, and Medicine, 2021], and as a top-three theme of the European Space Agency's (ESA's) *Voyage 2050* programme [European Space Agency *Voyage 2050*, 2021].

Exoplanet science is broadly moving from discovering planets to characterising them, including learning more about exoplanets in their stars' *habitable zone*, the distance at which the temperature allows for liquid water on a planetary surface. This shift is reflected in current and upcoming missions. While the trailblazing Kepler mission's goal was to determine how common planets are by observing one area of the sky [Borucki et al, 2010], upcoming decades will see the implementation of ELT (Extremely Large Telescope) missions with capabilities to determine the chemical makeups of exoplanetary atmospheres, focusing on bright host stars to capture the necessary detailed spectra [Kaltenegger, 2017]. The James Webb Space Telescope, launched in late 2021, uses transmission spectroscopy, comparing spectra of exoplanet host stars when the exoplanet is and is not transiting to determine which chemical species are present in the atmosphere [Beichman & Greene, 2018]. Early JWST data already provided evidence for water and clouds in the atmosphere of exoplanet WASP-96 b by summer 2022 [NASA "NASA's Webb", 2022]. ESA's planned upcoming PLATO and ARIEL space telescope missions will constrain masses, radii, densities, stellar irradiation, and ages of exoplanetary systems, and observe thermal structures and atmospheric compositions, respectively—all crucial pieces of information for assessing exoplanet habitability [Rauer et al, 2014; Tinetti et al, 2018]. These missions will bring this young fast-growing field closer to understanding exoplanets as whole worlds with unique environments.

The combination of recent discoveries in exoplanet science and the consistent presence of fictional exoplanets in SF form an ideal body of evidence for investigating the portrayal of science in SF. We are not aware of any comprehensive studies to date on exoplanet portrayal in SF—citations are usually sporadic and example-based, i.e. Chris Impey's popular science book *Worlds Without End* references Sagan's *Contact*, Liu's *Dark Forest*, Clarke's *The Fountains of Paradise* and Ackerman's poem "Ode to the Alien" [Impey, 2023]. Our research rectifies this gap, presenting a genre-wide corpus of fictional exoplanets to examine how the changing field of exoplanet science is presented in SF: how do variables like the story's medium, the physical characteristics of the fictional exoplanetary system, or the publication date affect the portrayal of science—and how do they relate to each other?

> *Research Question 2 (RQ2): Exoplanet science has made massive discoveries in recent decades. Are these findings reflected in science fiction exoplanets, and if so, how?*

*Bayesian Networks: Statistics & Literature*

To measure the influence of science on SF (RQ1) we focus on science fiction describing exoplanets (RQ2). There are precedents for using quantitative methods to study wider trends in genres, although traditional studies of literature tend to involve analysis of authors and characters and close reading—a technique focusing on a single story or story element. In the 1910s-20s, practitioners of the Formal Method argued for a more scientific study of literature that ignored the biographies and intentions of individual authors [Finer, 2010]. Instead, they focused on form, and on literature as a dataset to be examined in volume [Berlina, 2017]. Thus, Vladimir Propp's analysis of a body of folktales allowed him to identify 31 basic structural elements present in the genre and show his findings through diagrams and graphical networks [1968]. More recently, Franco Moretti's work uses 'distant reading' and graphics to chart changes in genres over time, uncovering structures within books [2005]. What these scholars share is that in losing the detail of individual works within their dataset they see the bigger picture, capturing wider trends in literature.

To answer RQ2, we need a big data approach to identify statistical dependencies within SF and changes in the body of literature over time. Moretti's methodologies borrow graphical structures from different fields to visualise literary data but do not do complex statistical analysis; Propp looks at structural elements within stories but not at generic trends. In this study, we create a Bayesian network to examine the interrelationships among features of fictional exoplanets and related scientific concepts, with the aim of determining the influence of scientific discovery on SF. Through our choice of variables (see Methodology), we will investigate how factors like media type, publication date, and fictional planetary system characteristics influence each other across a large dataset of SF. Bayesian networks are a versatile data science tool for trend mapping, and can aid in decision-making—this methodology has been applied to fields from genetics [Videla Rodriquez et al, 2022] to ecology [Hui et al, 2022] to astronomy [Pichara & Protopapas, 2013]. We are not aware of any literature discussing Bayesian networks applied to cultural or literary studies.

Bayesian networks allow us to analyse SF as a complex interconnected system. Given a dataset where each data point is a fictional exoplanet and its associated values for a set of variables (representing characteristics of the planet, such as atmosphere type), we can use Bayesian network analysis software to find a Bayesian network that describes the statistical dependencies in the dataset, i.e. how each variable influences the others. A Bayesian network is a graphical representation of a joint probability distribution—the probability of multiple events happening together—of a given set of variables [Heckerman et al, 1995]. Because it is not possible to simply calculate from a dataset the Bayesian network that describes its joint probability distribution [Chickering, 1996], unsupervised machine learning is used. In this instance a search algorithm explores potential sets of connections between variables and determines a network that best represents the statistical dependencies among measured variables in the data. These searches are typically done within a Bayesian statistical framework: in Bayesian statistics a 'prior' belief about a value is updated by data to generate a 'posterior' belief presented, alongside an estimate of its probability, as the solution. For a simple example: if using Bayesian statistics to calculate the average height of a group of people, one would start with a prior belief about height (e.g., adults are unlikely to be, e.g., 3 feet or 8 feet tall, and more likely to be within 5-7 feet), then measure the group in question and focus in on a range where, say, 95% of the heights lie. In a Bayesian network search, the *prior* represents a belief that any variables may be connected to each other, the *data* is your dataset, and the *posterior* is a network structure alongside a score representing how probable that structure is. A *search* uses these scores to find highly probable networks. A *search* starts with a random network and scores it against the measured dataset; it then makes a single change (adding or deleting a link), and asks if the score of this network is better or worse than the one before. By many such individual steps, millions of networks are explored and the structure of the Bayesian network that best fits the data is learned [Heckerman et al, 1995]. Bayesian networks come with arrows indicating a statistical directionality that can be read as 'is useful for predicting'; there can be multiple networks that are mathematically equivalent where the usefulness for predicting can go in either direction (e.g., clouds can predict it might rain, but rain also predicts existence of clouds). Bayesian networks provide an advantage over running simple statistics by enabling modelling of the entire context of measured variables, where subtleties of interactions among features of envisioned exoplanets can be placed within the larger context of the network and both direct (two variables linked in the network) and indirect (variables connected via a chain of links) connections can be explored.

> *Research Question 3 (RQ3): Bayesian networks are graphical representations of interconnected variables. What interactions between portrayals of fictional exoplanets and scientific knowledge can a network reveal?*

**Methodology & Results**

*Data:*

Our dataset is a collection of 142 fictional exoplanets from works of science fiction across different media. Planets were added to the database first from sources familiar to the authors, usually well-known SF franchises or classic books like *Star Wars, Star Trek, Dune, Solaris,* etcetera. To avoid bias towards only works we had encountered, we also used a crowdsourced Google form that collected fictional planet data from anonymous submissions. The form was shared in Twitter groups of astronomers and SF fans, and at conferences like the World Science Fiction Convention. Fictional exoplanets submitted were fact-checked and verified. Although there are SF universes with hundreds of fictional worlds, biases were mitigated by not including large numbers of worlds from the same property. The highest number of planets from a single source is eight from the *Star Trek* franchise.

While attempts were made to include planets from a variety of time periods and types of media, we recognise a current bias in the data toward Western, English-language fiction. To best probe RQ2, an effort was made to collect reasonably equal numbers

Variables for Fictional Exoplanet Database

| Variable Name | Description |
| --- | --- |
| AfterDiscovery | Whether the planet first appeared in fiction before (0) or after (1) the discovery of real-life exoplanets around sun-like stars in 1995. |
| HabZone | Whether the planet is shown to be in the liquid-water habitable zone (1) or outside it (0). |
| RealStar | Whether the planet is portrayed as being part of a real star system (1) or not (0). |
| Life | Whether the planet is home to native life (1) or not (0). |
| Intelligent | Whether the planet is home to intelligent native life (1) or not (0). |
| HumansBreathe | Whether human characters can breathe the planet's atmosphere without assistance or ill effect (1) or not (0). |
| MediaType | Whether the planet originally appeared in a film (0), a book (1), a TV show (2), a video game (3), or a podcast (4). |
| IsGas | Whether the planet is Earth-like and rocky/terrestrial (0) or Jupiter-like and gaseous (1). |
| EstNonNativeHumans | Whether the planet has been colonised by an established population of non-native humans who have been there for hundreds or thousands of years (1) or not (0). |

**Table 1:** Descriptions of the nine variables for the fictional exoplanet database, with numbers in parentheses indicating the value assigned to each state when reading the data into the Bayesian network software.

of fictional exoplanets from before and after 1995. Each fictional exoplanet was added to our database and categorised according to the nine variables in Table 1.

As an example, variables were assigned thusly for Tatooine:

*1 – AfterDiscovery:* 0. Tatooine first appeared in *Star Wars Episode IV: A New Hope* in 1977, before the discovery of real exoplanets [Lucas, 1977].
*2 – HabZone:* 1. Although Tatooine has a hot desert climate, its temperature range is temperate enough to support surface liquid water.
*3 – RealStar:* 0. *Star Wars* takes place 'in a galaxy far, far away'.
*4 – Life:* 1. Tatooine has native life such as sarlaccs and dewbacks.
*5 – Intelligent:* 1. In addition to these non-intelligent species, Tatooine is home to native intelligent species like Jawas and Tusken Raiders.

*6 – HumansBreathe:* 1. Human characters such as Luke Skywalker breathe the planet's atmosphere without assistance.
*7 – MediaType:* 0. Tatooine originally appeared in a film.
*8 – IsGas:* 0. Tatooine is a rocky, terrestrial world.
*9 – EstNonNativeHumans:* 1. There is an established population of non-native humans on Tatooine (in the *Star Wars* universe, humans are native to the planet Coruscant).

Variables were chosen to include scientifically-relevant statistics about the fictional exoplanets, with mostly binary options, since these aid in visually 'reading' the connections in the resulting network.

1995 was selected as the threshold year for AfterDiscovery because the first discovery of an exoplanet around a sunlike star, 51 Pegasi b, was made on October 6, 1995. Planets orbiting a pulsar, a remnant of a massive star after a supernova explosion, were discovered several years prior, but did not represent the same momentous proof of exoplanets around sunlike stars as the 1995 discovery [Wolszczan & Frail, 1992]. 1995 is widely accepted in the scientific community as the year exoplanets were discovered.

Some exoplanets in the database, such as Kepler-22b from *Raised by Wolves*, are fictional portrayals of real exoplanets. However, we did not include a variable for whether the planet is a fictional portrayal of a real planet because this would necessarily only occur after discovery of real exoplanets, and thus be a subset of the intersection of RealStar=1 and AfterDiscovery=1. We did not feel this would further our analysis as the author's scientific knowledge is already represented in RealStar. In contrast, Intelligent=1, despite being a subset of Life=1, provides additional details, showing (particularly with Life=1, Intelligent=0) understanding of the potential for extraterrestrial life.

While the concept of "habitable zones" did not enter wider use until after the discovery of exoplanets, we have included it because the concept long predates this discovery, first appearing in Maunder's *Are the Planets Inhabited?* [1913], and is easily retroactively applied to older works.

EstNonNativeHumans was a later addition to the list of variables. When determining whether life on a fictional exoplanet was native or non-native, many worlds were encountered that were home to intelligent life only in the form of non-native humans (especially common in large SF universes like *Star Wars* or *Foundation* or the *Hainish Cycle*). This occurred often enough to merit the addition of its own category.

We plotted a bar graph (Figure 1) for the 142 planets in the database, showing the distribution of categories for each of the 9 variables as percentages. The bar graph allows us to check for representation in our database.

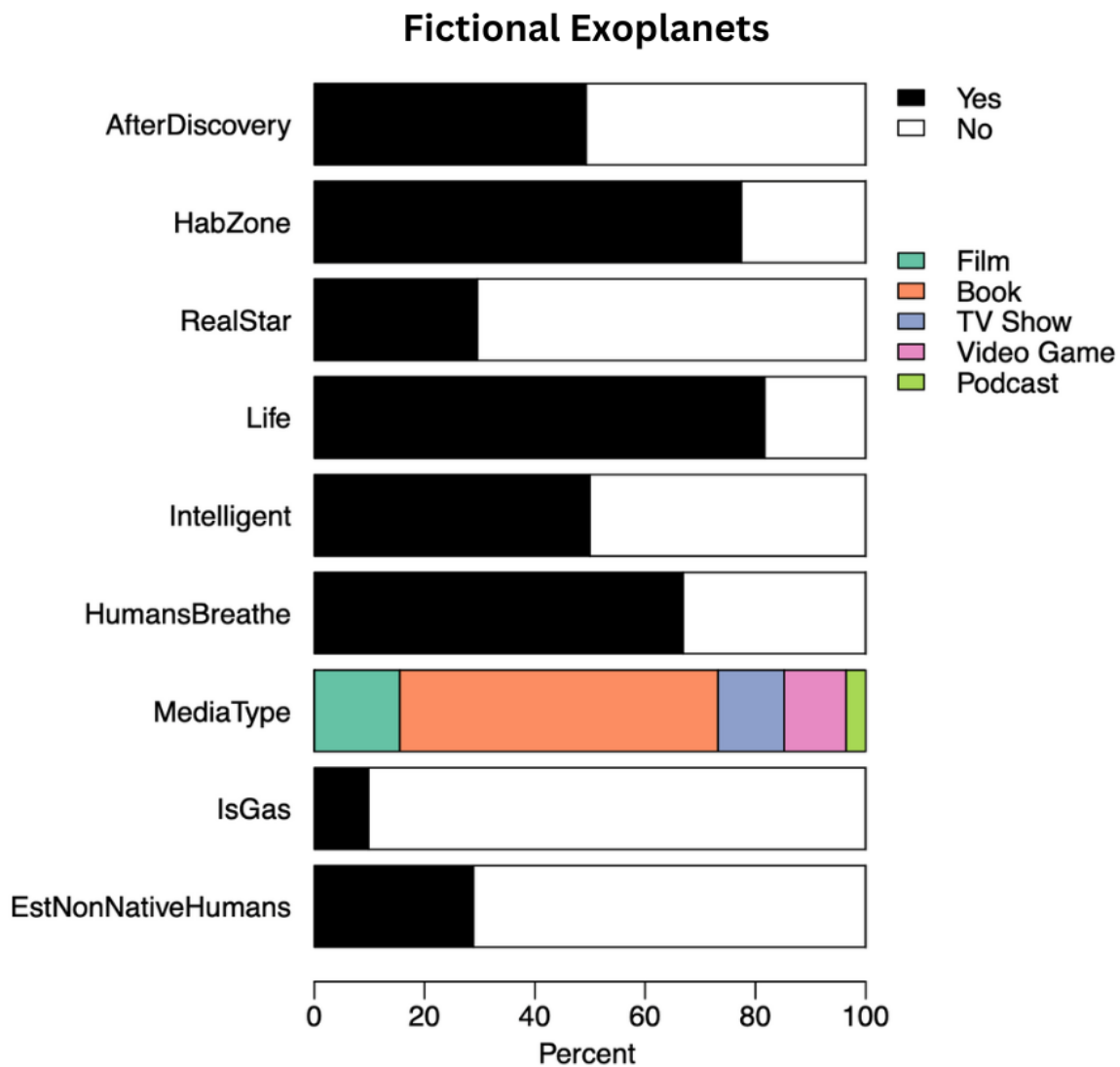

**Figure 1:** Distribution of categories for each variable in the fictional exoplanet database. For each variable, a bar is filled to show the percentage of planets which took on each categorical value: Yes (coded as 1), No (0) or the types of media for MediaType. The larger proportions of "Yes" for HabZone, Life, HumansBreathe and "No" for IsGas show the majority of worlds had Earth-like features.

There is a near-even split of fictional exoplanets from before and after the discovery of real exoplanets, showing our attempt to collect equal amounts of examples for each was successful. Many planets have Earth-like features—more than half the fictional exoplanets are located in the habitable zone, are places where human characters can breathe, and are home to native life—with half home to intelligent native life. 62% of the planets in the database are terrestrial worlds in the habitable zone with human-breathable air. About 30% orbit real stars, and the same number are home to established non-native human populations. Only 10% of the fictional exoplanets in the database are gas giants. While more than half of our fictional exoplanets come from literature, we have significant representation from video games, film, television, and podcasts.

*Network*

To find a Bayesian network that represents the statistical dependencies between the nine variables within the fictional exoplanet database, we used the software Banjo [Hartemink, 2010] to generate a network structure along with an influence score for each link, which measures the strength and direction of the relationship between variables. Influence scores can be positive or negative, indicating direction, with strength indicated by distance from zero. Influence scores can also be reported as non-monotonic, meaning there is no specific direction to the relationship, e.g., where a high value of one variable maps to both high and low values of another [Yu et al, 2004].

Banjo uses the Bayesian Dirichlet equivalent (BDe) scoring metric to score the networks and can produce as a solution either a top-scoring network or an average of a user-defined number of top-scoring networks. Banjo supports two types of search algorithm, *greedy* and *simulated annealing*, and within these two ways of making individual steps between networks, and further allows the user to define the period of time a search takes (longer searches explore more networks). *Greedy* search always picks the higher-scoring option when changing links to try a new network [Jungnickel, 1999]; this means it always climbs to the nearest 'peak'—a network where all networks one step away score lower—from its random starting point. *Simulated annealing* search has a decreasing probability of choosing the lower-scoring option [Kirkpatrick et al, 1983]. *Simulated annealing* can take a longer time to find high-scoring networks than *greedy*, but can also find high-scoring networks surrounded by lower peaks, whereas *greedy* would stop at the low peak and never go down and back up to find the higher one. When taking individual steps in the search, Banjo allows users to choose either *random local moves* (makes random changes in network links and takes the first better-scoring option) or *all local moves* (checks all possible changes and goes to the option which most improves the score). *Random local moves* can enable a broader search of structure, as it goes through more different networks faster, but *all local moves* ensures that a nearby higher-scoring structure is not missed.

Given all these trade-offs in the user-defined settings, we varied various parameters as described below to determine the best solution to use. We ran 48 total searches in Banjo, in which a *search* means one instance of the algorithm creating many possible networks in turn and scoring them, using a particular combination of settings. We explored (a) *greedy* and *simulated annealing* search, (b) *random local moves* and *all local moves*, and (c) allowed the search to run for 1, 5, 10, and 15 minutes, for a total of 2x2x4=16 different parameter combinations, and we repeated each combination for 3 searches.

The only other parameters that need to be set, and which we did not vary, are the *equivalent sample size* and *best networks are*. The equivalent sample size reflects the level of weight to give to a Bayesian prior—a belief that any given link is present—in relative terms to data points (here, exoplanets); a low value so that your data provides the bulk of knowledge is generally used: we picked 1. The best networks can be set to either allow equivalent networks, or to ensure that all networks reported represent different statistical descriptions of the data; the latter is better and only avoided for computing time reasons: we had no issues with computing time thus set the best networks to be nonequivalent.

All 48 network searches resulted in the same highest-scoring network. We visualised the searches in the software BayesPiles [Vogogias et al, 2018] and saw that many different search paths all resulted in the same top network. This is a clear signal that this single top network should be used as a solution. Our a highest-scoring network had nine links connecting eight of the nine variables, and is displayed in Figure 2.

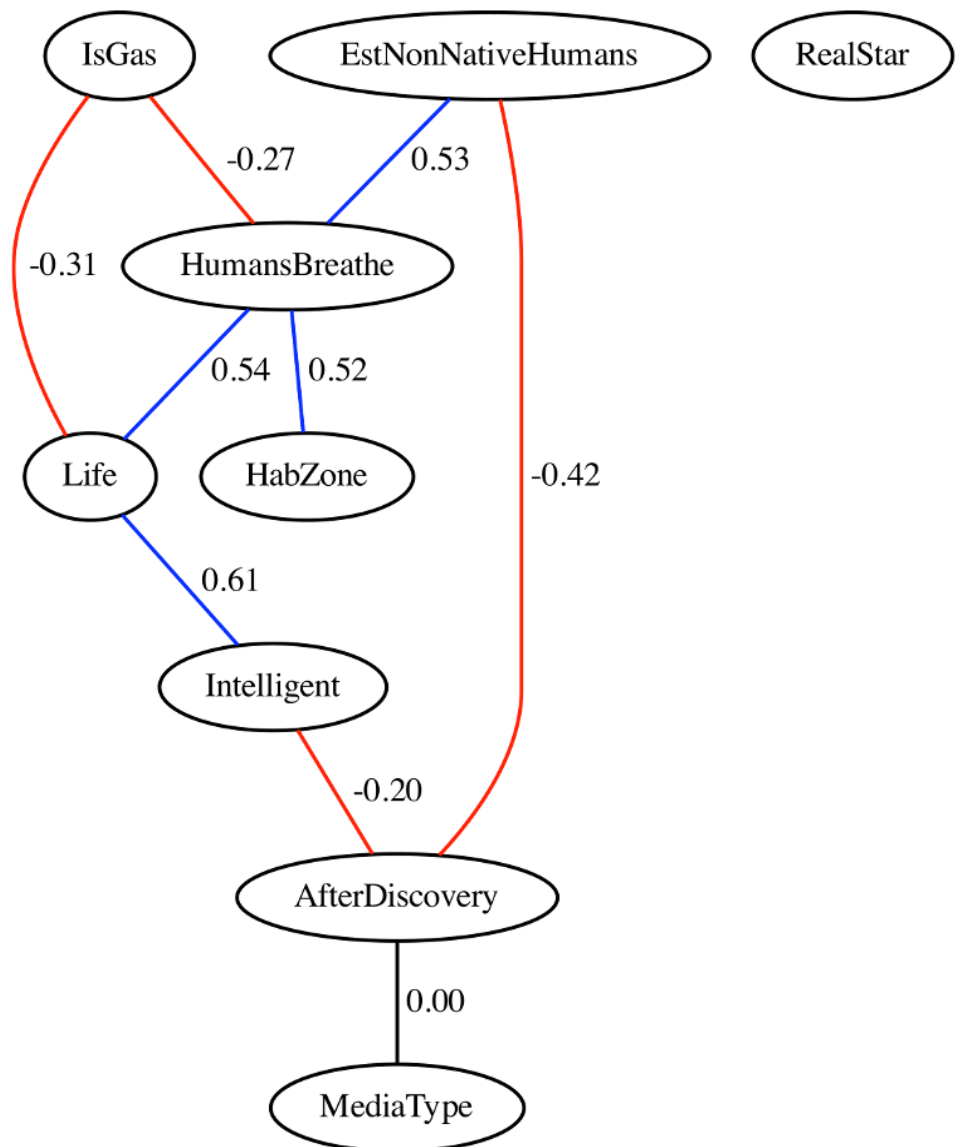

**Figure 2:** The highest-scoring network. Blue links represent positive influence, red represent negative, and black non-monotonic (see above). Variables are encased in ovals; influence scores are next to their links. While Bayesian networks come with directed links (arrows), we show here the links without direction: mathematically equivalent networks can have arrows going in either direction and as we searched for nonequivalent networks only we show the links without direction; further, we are not claiming causality but instead informativeness, and an undirected interaction better reflects this property.

To summarise, our Bayesian network contains nine links describing how fictional exoplanet characteristics influence each other. The five variables with positive links between them—Life, Intelligent, HabZone, EstNonNativeHumans, and HumansBreathe—form a cluster. These features being true indicates a more Earth-like planet, showing the interconnectedness of these variables in worldbuilding considerations. There are negative influence links between IsGas and HumansBreathe and IsGas and Life, showing gaseous planets are negatively associated with these Earth-like features. There are also negative influence links between EstNonNativeHumans and AfterDiscovery and Intelligent and AfterDiscovery, showing that fictional planets with established non-native humans or with intelligent native life were less common after the real-life discovery of exoplanets. These negative links connect AfterDiscovery indirectly to the positive cluster of Earth-like links, showing that of that cluster, EstNonNativeHumans and Intelligent are the two Earth-like traits likeliest to be absent in a post-detection fictional exoplanet. The network thus demonstrates that the changing likelihood of Earth-like fictional exoplanets before and after the discovery of real exoplanets is mediated by the presence or absence of intelligent life—be it native or in the form of non-native humans. There is a non-monotonic link between AfterDiscovery and our only nonbinary variable, MediaType. As described above, this is a link too complicated to show in this Figure due to MediaType's having five states. More recent SF stories are presented in a wider variety of media, and this link is caused by the increased prevalence of planets sourced from video games and podcasts in the post-1995 data. All variables except RealStar have at least one link to another variable, showing an absence of influence of whether or not the planet setting is in a real star system on other worldbuilding characteristics. The Bayesian network thus provides an overview of the landscape of science fiction exoplanets within our dataset.

**Discussion**

To answer RQ1, we are interested in what the network reveals about the portrayal of science in SF. Results regarding AfterDiscovery represent what has changed in the process of imagining fictional exoplanets in the years since the real-life discovery of exoplanets. The network shows AfterDiscovery negatively influences Intelligent and EstNonNativeHumans, meaning that there have been fewer portrayals of intelligent native life and established non-native human populations on fictional exoplanets created since 1995. These negative links connect AfterDiscovery to the positive Earth-like cluster, showing a decrease in Earth-like characteristics post-1995. RQ2 seeks to understand if exoplanet science findings are reflected in SF, and indeed, this decrease in Earth-like fictional exoplanets matches scientific findings. With the discovery of real exoplanets and the revelation that very un-Earth-like worlds like hot Jupiters or tidally-locked Super-Earths might be common, humanity has uncovered a variety of exotic worlds that are unlikely to be hospitable to humans. Popular news media reports on exoplanet discoveries, which would be accessible to SF authors, highlight hot Jupiters trailing comet tails of atmosphere [Choi, 2015], worlds that rain diamonds [Strickland, 2022], and worlds that orbit multiple stars [Carter, 2022]. With increasing cultural awareness of extreme exoplanets and of the concept of a habitable zone, creators may have been inspired to explore what stranger new worlds outside this zone might be like. Our Bayesian network reveals changes in SF portrayal of exoplanet science with new discoveries, addressing RQ3.

Bayesian network analysis provides a statistical methodology with which to uncover trends in SF portrayal of exoplanets on a larger scale than traditional example-based close reading. The network can be used as an overview of the complete dataset, but it is also a starting place to look at detail in how science is portrayed in SF. The network highlights that these AfterDiscovery links are mediated by the variables associated with intelligent life (Intelligent and EstNonNativeHumans). There are 44 planets in the database that have 1 for Life and 0 for Intelligent—that is, planets that host non-intelligent life. Of those planets that have non-intelligent native life but no intelligent native life, many of those from before the discovery of real exoplanets also have 1s for EstNonNativeHumans. Therefore, while intelligent life is depicted as living and thriving there, it is intelligent life that is not native to the planet. Examples of unintelligent native life and established non-native humans include older works like Urras and Anarres from Ursula Le Guin's *The Dispossessed*, and Arrakis from Frank Herbert's *Dune*. This combination of characteristics is not gone from more recent works – Nilt from Ann Leckie's *Ancillary Justice* is an example. However, planets with *only* non-intelligent life are more common in recent works: M6-117 from *The Chronicles of Riddick*, LV-1201 from *Aliens versus Predator 2*, and all four worlds from Becky Chambers' *To Be Taught, If Fortunate* serve as examples. The Bayesian network thus illustrates, through its links between AfterDiscovery and Intelligent and EstNonNativeHumans, the story of how more recent science fiction is more likely to feature planets with un-intelligent native biospheres, addressing all three RQs.

The strength of the Earth-like cluster in the network demonstrates the frequency of planets with Earth-like characteristics (life, rocky surface, human-breathable atmosphere, intelligent life) in SF. In light of RQ1 about the portrayal of science in SF, this finding highlights an important consideration for any practitioners looking to incorporate SF into their science communication work: that the population of fictional exoplanets still looks very different from the population of real exoplanets. Given the youth of the field of exoplanet science, we are only just beginning to perform population statistics on discovered exoplanets and extrapolate about the overall population, of which the 5000-odd exoplanets discovered constitute a tiny fraction. Still, most discovered exoplanets are not human-habitable. Of the 4000 planets in the Kepler confirmed and candidate population, only about 7% orbit in the habitable zone, as defined by placing limits on solar insolation received and equilibrium temperature [NASA Exoplanet Science Institute, 2023], and being in the habitable zone says nothing about, for example, a planet's atmosphere. Our fictional exoplanets database, in contrast, has more than half its worlds in the habitable zone, and the network reveals strong links between the variables that indicate Earth-like conditions. The dataset of real exoplanets contains far fewer Earth-like worlds as a percentage than our fictional dataset, often in practice driven by narrative purposes in telling stories about humans. Applying Bayesian network analysis reveals the Earth-like natures of the fictional exoplanet population.

Our Bayesian network allows us to see the big-picture, interconnected view of fictional exoplanet worldbuilding, and to make informed proposals for explanations for the links. While SF exoplanets have an Earth-like bias, they still explore physical and astrobiological scientific concepts. As we've shown, the majority of fictional exoplanets host native extra-terrestrial life. Alien environments might not interact well with human biological needs. Exoplanet discoveries, combined with a larger shift in understanding of environmental science and the rise of eco-anxiety [Coffey et al, 2021], have led to an increasing awareness of the interconnectedness of systems, and the ecological damage that could be wrought by separating biology from its own planet may be reflected in the changes shown by the network in recent SF. Human settlers on the titular planet from *Aurora* die of prion disease from their new exoplanet home [Robinson, 2015], and human settlers on Laconia in *The Expanse* cannot eat the native food on their new world because life there exhibits mirror chirality [Abraham & Woolnough, 2021]. This new awareness may be behind the decrease in established non-native human populations on fictional exoplanets in recent works. Taken in combination, our findings of the Earth-like cluster and its negative links with AfterDiscovery, as well as the lack of links involving RealStar, suggest that an author/creator's level of scientific knowledge is not reflected directly in the features of exoplanets imagined. Instead, these features are more connected to the gestalt presented by the overall community of knowledge, a big-picture finding made possible by the Bayesian network methodology. That fictional exoplanets are becoming less human-habitable is a demonstration of exoplanet science findings within science fiction, with the Bayesian network revealing the interconnected variables at play.

Our results show that SF is responsive to changes in scientific knowledge. Trends learned from Bayesian network analysis could be used to guide investigations into the corpus of science fiction media for purposes of science communication and outreach, by helping select fictional exoplanets with characteristics of interest for teaching, thus mitigating concerns about SF misrepresentations of science.

It should be noted that our study has some limitations. Our network should be considered in context of its English-language bias, and by the need to fill every category for every fictional exoplanet. While theory exists for handling missing data in Bayesian network structure learning [Friedman, 1998], there is a lack of software which provides sufficient implementation and interoperability of this with other tools for gaining confidence in resulting network structures. Simple imputation methods, such as replacing empty values with the most frequently occurring value, would skew our data too much. Therefore, we only included a planet in the database if we could answer every variable, thus excluding worlds only briefly visited in their stories.

**Conclusions**

While scientists sometimes use SF for science communication, and there is increased interest in this among scholars, there have been concerns about scientific misconceptions within SF and few studies have investigated the science content of SF. Our Bayesian network of the fictional exoplanet database begins to address this gap by statistically evidencing the idea, previously reported through case studies and close reading, that SF changes in response to scientific discovery, incorporating scientific results into its stories. Our Discussion demonstrates that the Bayesian network results reveal trends in science fiction exoplanets that can guide further research and SF use in science communication. These results can be augmented with specific examples from SF, allowing for more scrupulous and informed close reading and fiction analysis and providing an example of how the network results could aid interested science communicators in making informed uses of SF in their work with the public. While our findings point towards strong potential for a role for SF in science communication and provide an adaptable methodology for monitoring science content in SF, further research could develop methodologies for delineating fact from fiction, or developing curricula to teach science using SF.

Bayesian network analysis has the power to uncover larger trends within the genre of science fiction, and from other digital humanities and media studies datasets, that are impossible to learn from close readings alone. This methodology is highly applicable to different problem sets, including helping science communication practitioners understand media portrayal of other fields of science besides exoplanets. We provided detail on the theory behind Bayesian networks in Context and Methods, and highlight that science communicators need not be statisticians to be users of this methodology: informative and accessible tutorial courses can be found at B-course [Myllymäki et al, 2002], and Banjo, the software we used, provides extensive documentation to aid the new user [Hartemink, 2010]. Through our application of this methodology, we reveal changes in SF media towards less Earth-like and human-habitable portrayals of fictional exoplanets. Our results show that science fiction as a genre does evolve alongside scientific discovery, reflecting the findings of the growing field of exoplanet science. Science fiction, though previously understudied as a means of science communication, is attracting more interest as a teaching tool due to its large audience and public interest. We argue that understanding the genre's portrayal of science is essential for its development as a science communication tool, and provide quantitative rigour to this endeavour with an innovative data science methodology.

**Data Availability Statement**
*The research data underpinning this publication can be accessed at <https://doi.org/10.17630/5a15a8c8-683b-421f-a24d-b68b6d5f1c0e>.*

**Authors**

Emma Johanna Puranen is a St Leonards' Interdisciplinary Doctoral Scholar at the St Andrews Centre for Exoplanet Science. Her research bridges astronomy, statistics, and media studies to turn science fiction into data science. She also publishes on space ethics and humanity's future in space, and is a science fiction author whose work has been reviewed in Nature Astronomy.

Emily Finer is senior lecturer at the University of St Andrews in Scotland and co-director of the interdisciplinary St Andrews Centre for Exoplanet Science. She publishes on translation and translanguaging between Russian, English, Yiddish, and Polish, and her research focuses on literature, culture, and science in transnational contexts. She is Principal Investigator for a collaborative interdisciplinary research project: "Forecasting Reproduction in Space".

Christiane Helling habilitated in Astrophysics at the TU Berlin (DE). After several PostDoc positions, she became lecturer and then reader at the University of St Andrews (UK). Later, she became the director of the St Andrews Centre for Exoplanet Science and, in addition, full professor. She then took over the positions as director of the Space Research Institute in Graz (AT) and full professor for space science at the TU Graz. She has (co-)authored more than 135 refereed publications.

V. Anne Smith is a senior lecturer in the School of Biology and associate dean curriculum for the faculty of Science at the University of St Andrews. Her research interests focus on the development and application of computational models to understand complex biological/biosocial systems, ranging from gene regulation to behaviour. She is author of A Code For Carolyn: A Genomic Thriller, which forms part of Springer's Science and Fiction series.